\documentclass[12pt]{iopart}
\usepackage{iopams}
\usepackage{amsfonts}
\usepackage{graphicx}
\usepackage{subfigure}
\usepackage{amsthm}
\usepackage{times}
\usepackage{color}
\usepackage{ulem}
\usepackage[compress]{cite}

\newcommand{\eqref}[1]{{(\ref{#1})}}
\begin{document}


\title[]{Algebraic structure of the two-qubit quantum Rabi model and its solvability using Bogoliubov operators}

\author{Jie Peng$^{1,2}$, Zhongzhou Ren$^{2,3,4,5}$, Haitao Yang$^6$, Guangjie Guo$^{7}$, Xin Zhang$^{2,3}$, Guoxing Ju$^{2}$, Xiaoyong Guo$^{8}$, Chaosheng Deng$^{1}$, Guolin Hao$^{1}$}
\address{$^{1}$Laboratory for Quantum Engineering and Micro-Nano Energy Technology and School of Physics and Optoelectronics, Xiangtan University, Hunan 411105, China}
\address{$^{2}$Key Laboratory of Modern Acoustics and Department of Physics, Nanjing University, Nanjing 210093, China}
\address{$^{3}$Joint Center of Nuclear Science and Technology, Nanjing University, Nanjing 210093, China}
\address{$^{4}$Center of Theoretical Nuclear Physics, National  Laboratory of Heavy-Ion Accelerator, Lanzhou 730000, China}
\address{$^{5}$Kavli Institute for Theoretical Physics China, Beijing 100190, China}
\address{$^{6}$College of physics and electronic information engineering, Zhaotong University, Zhaotong 657000, China}
\address{$^{7}$Department of Physics, Xingtai University, Xingtai 054001, China}
\address{$^{8}$School of Science, Tianjin University of Science and Technology, Tianjin 300457, China}

\ead{\mailto{pengjie145@163.com}, \mailto{zren@nju.edu.cn} and
\mailto{jugx@nju.edu.cn}}

\date{\today}

\begin{abstract}
We have found the algebraic structure of the two-qubit quantum Rabi
model behind the possibility of its novel quasi-exact solutions with
finite photon numbers by analyzing the Hamiltonian in the photon
number space. The quasi-exact eigenstates with at most $1$ photon
exist in the whole qubit-photon coupling regime with constant
eigenenergy equal to single photon energy $\hbar\omega$, which can
be clear demonstrated from the Hamiltonian structure. With similar
method, we find these special ``dark states''-like eigenstates
commonly exist for the two-qubit Jaynes-Cummings model, with
$E=N\hbar\omega$ ($N=-1,0,1,\ldots$), and one of them is also the
eigenstate of the two-qubit quantum Rabi model, which may provide
some interesting application in a simper way. Besides, using
Bogoliubov operators, we analytically retrieve the solution of the
general two-qubit quantum Rabi model. In this more concise and
physical way, without using Bargmann space, we clearly see how the
eigenvalues of the infinite-dimensional two-qubit quantum Rabi
Hamiltonian are determined by convergent power series, so that the
solution can reach arbitrary accuracy reasonably because of the
convergence property.
\end{abstract}


\maketitle

\section{Introduction}
\label{intro} \noindent The quantum Rabi model \cite{jc} describes
the interaction between a bosonic mode and a two--level
system---probably the simplest interaction between light and matter.
Its semiclassical form was first introduced by Rabi in nuclear
magnetic resonance \cite{rabi}. In 1963, Jaynes and Cummings
\cite{jc} found its application in describing the interaction
between a two-level molecular and a single mode photon field. With
the developments of experiments, many systems can be described by
this model in quantum optics \cite{guo}, condensed matter
\cite{irish}, cavity quantum electrodynamics (QED) \cite{arne},
circuit QED \cite{ad}, quantum dots \cite{zhong}, trapped ions
\cite{trapped} and so on. Although this model takes a very simple
form, its analytical solution was not so easy to obtain, so various
approximations were made, one of which is the famous
``rotating--wave approximation'' \cite{jc}. In 2011, its solution
was analytically found by Braak \cite{br} in the Bargmann space
\cite{barg}. It can describe the ultrastrong qubit-photon coupling
regime, which has been reached in recent circuit QED experiments
\cite{nie}, where the ``rotating wave approximation'' breaks down.
After that, various researches are done to the full Rabi
Hamiltonian, including recovering the solution of the Rabi model
\cite{qing,braak,bra2}, real-time dynamics \cite{bra1}, the solution
of the two-qubit Rabi Hamiltonian
\cite{pj,2bite,pj1,pj2,qinghuchen}, dynamical correlation functions
\cite{bra2a}, and so on
\cite{trave,braak2,yunbo,yaozhong,moro,chen1,chen2,nanjing,bd,ff}.

Two-qubit system is basic and fundamental to the construction of the
universal quantum gate. Various qubit-qubit interactions are applied
to generate qubit-qubit entanglement and realize quantum computation
\cite{gea,you}, one of which is mediated by a resonant cavity,
described by the two-qubit quantum Rabi model \cite{pj2}. In this
case, the ultrafast two-qubit quantum gate can be constructed in the
ultrafast qubit-photon coupling regime \cite{rg}. Besides, the
distant qubits can be coupled through a resonant cavity and the
coherent quantum state storage and transfer can be realized
\cite{cp1}. Working for the whole qubit-photon coupling regime, the
two-qubit quantum Rabi model can be applied in many systems in
quantum optics \cite{fa} and quantum information \cite{zrhk}. Its
analytical solution was obtained in \cite{pj2} by means of Bargmann
space approach, and also in \cite{qinghuchen} with extended coherent
states representation. One interesting result is that there exist
coupling-dependent eigenstates in the whole coupling regime with
constant eigenenergy--reminiscent of ``dark states'', but they are
coupling-dependent and the photon number is bounded from above at
$1$, which is novel and interesting. Besides, there are quasi-exact
solutions with finite photon numbers $N$, which are not presented in
the one-qubit Rabi model. These special solutions may have some
interesting application, however, the algebraic structure behind the
possibility of these special solutions needs to be clarified.

In this paper, we have clarified the algebraic structure of the
two-qubit quantum Rabi model for its special quasi-exact solutions
with finite photon numbers found in \cite{pj2}. By analyzing the
Hamiltonian in the photon number space, we find the condition for
closed subspace, i.e. the algebraic structure are related with the
permutation symmetry of the qubit-photon coupling terms for the two
qubits. Even more interestingly, the quasi-exact solution with at
most $1$ photon exists in the whole coupling regime with constant
eigenenergy equal to single photon energy $\hbar\omega$, which can
be clearly found from the algebraic structure. These eigenstates are
partly like ``dark states'', but are coupling dependent and the
photon number is bounded from above, so they may have some
interesting. According to the algebraic structure of the two-qubit
quantum Rabi model, we may conjecture there are similar ``dark
states''--like solutions to those models with homogenous
qubits-photon coupling terms. For example, we consider the two-qubit
Jaynes-Cummings model \cite{cp1}, which is commonly applied for
simplicity in the weak coupling regime \cite{2jc}. Very
interestingly, under similar condition, we find many ``dark
states''--like eigenstates, existing in the whole coupling regime
with constant eigenenergy $E=N\hbar\omega~(N=-1,0,1,\ldots)$, one of
which is also the eigenstate of the two-qubit Rabi model. Since the
Jaynes-Cumming model is simper than the Rabi model, these
eigenstates may provide some interesting application easier. On the
other hand, we analytically retrieve the solution of the two-qubit
quantum Rabi model, using Bogoliubov operators. With this more
physical and straightforward method, we find a way to obtain its
solution by convergent power series, so that we can make reasonable
cutoff in practical calculation and the solution can reach arbitrary
accuracy.

The paper is organized as follows.  In section \ref{s2}, we clarify
the algebraic structure behind the possibility of quasi-exact
solutions with finite photon numbers obtained in \cite{pj2} and also
find the special ``dark states''--like solutions of the two-qubit
Jaynes-Cummings model. In section \ref{s3}, we analytically retrieve
the solution of the two-qubit quantum Rabi model using Bogoliubov
operators. Finally, we make some conclusions in section \ref{s4}.

\section{Algebraic structure for quasi-exact solutions with finite photon numbers}\label{s2}
The Hamiltonian of the two-qubit quantum Rabi model reads
($\hbar=1$) \cite{2bite,pj2}
\begin{equation}\label{gq}
H_{tq}=\omega
a^{\dagger}a+g_{1}\sigma_{1x}(a+a^{\dagger})+g_{2}\sigma_{2x}(a+a^{\dagger})
+\Delta_1\sigma_{1z}+\Delta_2\sigma_{2z},
\end{equation}
where $a^{\dagger}$ and $a$ are the single mode photon creation and
annihilation operators with frequency $\omega$, respectively.
$\sigma_{i}\, (i=x,y,z)$ are the Pauli matrices. $2\Delta_1$,
$2\Delta_2$ are the energy level splittings of the two qubits.
$g_{1}$ and $g_{2}$ are the qubit-photon coupling constants for the
two qubits respectively. There are quasi-exact solutions with finite
photon numbers $N$ obtained by analyzing the recurrence relation of
the coefficients in \cite{pj2}. However, the algebraic structure
behind the possibility of these novel exceptional solutions needs to
be clarified.

Quasi-exact solutions with finite photon numbers $N$ correspond to
the existence of closed subspace in the photon number
representation, i.e. the algebraic structure. Here we demonstrate
the closed subspace are related with the permutation symmetry of the
qubit-photon coupling terms by analyzing the structure of the
Hamiltonian in the photon number space. $H_{tq}$ \eqref{gq} process
a $\mathbb{Z}_2$ symmetry with the transformation $R=\exp(i\pi
a^\dagger a)\otimes\sigma_{1z}\otimes\sigma_{2z}$. Taking odd parity
for example, supposing the initial state $|\psi\rangle$ is in a
subspace formed by $\{|M,e,g\rangle,|M,g,e\rangle,|M+1,g,
g\rangle,|M+1,e,e\rangle,\cdots,|N-1,g,g\rangle,|N-1,e,e\rangle,
|N,e,g\rangle,\\|N,g,e\rangle\}$, with the coefficient
$\{\large{c}_{1,M},\large{c}_{2,M},\large{c}_{1,M+1},\large{c}_{2,M+1},\ldots,\large{c}_{1,N},\large{c}_{2,N}\}$,
where $M$ and $N$ are even, then the Hamiltonian reads ($\omega$ is
set to $1$)
\begin{eqnarray}\small\fl
\left(
  \begin{array}{ccccccc}
   0 & 0&\sqrt{M}g_1 & \sqrt{M}g_2 & 0&0&\dots  \\
    0 & 0&\sqrt{M}g_2 & \sqrt{M}g_1 & 0&0&\dots  \\
   \sqrt{M}g_1&\sqrt{M}g_2& M+\Delta_1-\Delta_2 & 0 &\sqrt{M+1}g_1&\sqrt{M+1}g_2 &\dots \\
   \sqrt{M}g_2&\sqrt{M}g_1& 0&M+\Delta_2-\Delta_1 &\sqrt{M+1}g_2&\sqrt{M+1}g_1
   &\dots\\
   \dots&\dots&\dots&\dots&\dots&\dots&\dots
   \end{array}
   \right)\nonumber\\
   \small\fl
   \left(
    \begin{array}{ccccccc}
     \dots&\dots&\dots&\dots&\dots&\dots&\dots\\
    \dots & \sqrt{N}g_1 & \sqrt{N}g_2 & N+\Delta_1-\Delta_2 & 0 & \sqrt{N+1}g_1 & \sqrt{N+1}g_2 \\
    \dots & \sqrt{N}g_2 & \sqrt{N}g_1 & 0 & N+\Delta_2-\Delta_1 & \sqrt{N+1}g_2& \sqrt{N+1}g_1 \\
   \dots& 0&0 & \sqrt{N+1}g_1 & \sqrt{N+1}g_2 & 0 & 0 \\
    \dots& 0&0  & \sqrt{N+1}g_2 & \sqrt{N+1}g_1 & 0 & 0
  \end{array}
\right).
\end{eqnarray}
If for $|\psi^\prime\rangle=H|\psi\rangle$, the coefficients of
$\{|N+1,g,g\rangle,~|N+1,e,e\rangle\}$ and
$\{|M-1,g,g\rangle,~|M-1,e,e\rangle\}$ equal to $0$, then this
subspace is closed. For the first case, we obtain
\begin{eqnarray}
\sqrt{N+1}\large{g}_{\scriptstyle 1} \large{c}_{1,N}+\sqrt{N+1}g_2 c_{2,N}=0,\label{n1}\\
\sqrt{N+1} g_2 c_{1,N}+\sqrt{N+1}g_1 c_{2,N}=0, \label{n2}
\end{eqnarray}
where $c_{1,N}$ and $c_{2,N}$ are the coefficients of
$|N,e,g\rangle$ and $|N,g,e\rangle$ respectively. From equations
\eqref{n1} and \eqref{n2} and $g_1,~g_2>0$, we obtain $g_1=g_2$ and
$c_{1,N}=-c_{2,N}$. By using the time-independent Sch\"{o}dinger
equation, we obtain
\begin{eqnarray}
\sqrt{N}g_1 c_{1,N-1} + \sqrt{N}g_2 c_ {2,N-1} +
(N+\Delta_1-\Delta_2)c_{1,N}=Ec_{1,N},\\
\sqrt{N}g_2 c_{1,N-1} + \sqrt{N}g_1 c_{2,N-1} +
(N+\Delta_2-\Delta_1)c_{2,N}=Ec_{2,N}, \end{eqnarray} so that
\begin{eqnarray}
E=N,\\
(\Delta_2-\Delta_1)c_{1,N}=(\sqrt{N}g_1 c_{1,N-1} + \sqrt{N}g_2
c_{2,N-1}).
\end{eqnarray}
For the special case $\Delta_1=\Delta_2$ and
$c_{1,N-1}=c_{2,N-1}=0$, there is a invariant subspace formed by
$\{|N,e,g\rangle,|N,g,e\rangle\}$, and the eigenstate is
\begin{equation}
|\psi\rangle_{N}=\frac{1}{\sqrt{2}}(|N,g,e\rangle-|N,e,g\rangle),
\end{equation}
which is the famous ``dark state'' \cite{rod-lara,pj2}, where the
spin singlet is decoupled from the photon field.

If $\Delta_1\neq\Delta_2$, considering the coefficient of
$\{|M-1,g,g\rangle,~|M-1,e,e\rangle\}$ must be $0$, it is required
that $E=M$, which contradicts with $E=N$, so that the only possible
choice is $M=0$. Now we have obtained a closed subspace (algebraic
structure) formed by $\{|0,e,g\rangle,|0,g,e\rangle,|1,g,
g\rangle,|1,e,e\rangle,\cdots,|N-1,g,g\rangle,|N-1,e,e\rangle,
|N,e,g\rangle,|N,g,e\rangle\}$, with the condition
\begin{eqnarray}
g_1=g_2,\\
E=N. \end{eqnarray} Then by using the time-independent
Sch\"{o}dinger equation, we can obtain quasi-exact solutions with
finite photon number $N$ for certain choice of parameters
$\Delta_1$,~$\Delta_2$,~and $g=g_1+g_2$. For example, if $N=2$, the
determinant of the matrix
\begin{eqnarray}\small\fl
 \left(
  \begin{array}{cccccc}
    \Delta_1-\Delta_2-2 & 0                     & g/2                     & g/2                  & 0                   & 0  \\
    0                   & \Delta_2-\Delta_1-2   & g/2                     & g/2                  & 0                   & 0  \\
    g/2                 & g/2                   & -\Delta_1-\Delta_2-1    & 0                    & \sqrt{2}g/2         & \sqrt{2}g/2  \\
    g/2                 & g/2                   & 0                       & \Delta_2+\Delta_1-1  & \sqrt{2}g/2         & \sqrt{2}g/2  \\
    0                   & 0                     & \sqrt{2}g/2             & \sqrt{2}g/2          & \Delta_1-\Delta_2   & 0  \\
    0                   & 0                     & \sqrt{2}g/2             & \sqrt{2}g/2          & 0                   &\Delta_2-\Delta_1   \\
   \end{array}
   \right)
   \end{eqnarray}
   must equal to $0$, which gives
\begin{equation}
(\Delta_1^2-\Delta_2^2)[(4-(\Delta_1-\Delta_2)^2)(1-(\Delta_1+\Delta_2)^2)-2g^2]=0.
\end{equation}

\begin{figure}[htbp]
\center
\includegraphics[width=0.8\textwidth]{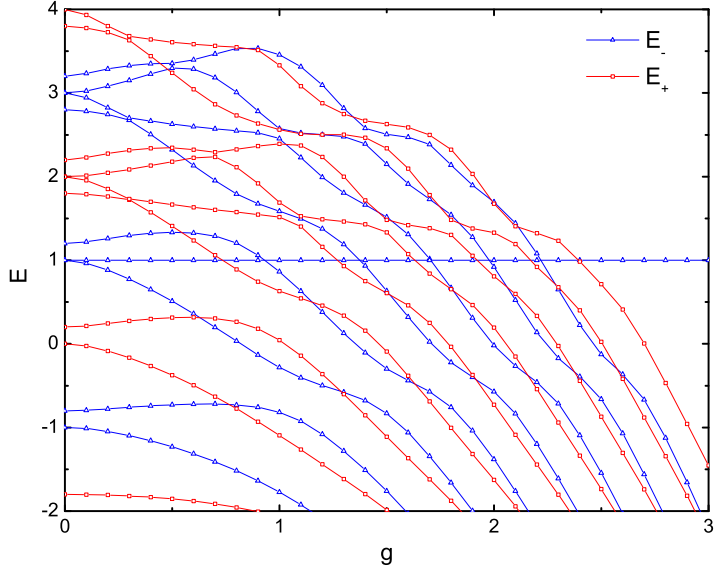}
\renewcommand\figurename{\textbf{Figure}}
\caption[2]{The numerical spectrum of two-qubit quantum Rabi model
with $\Delta_1=1.4$,~$\Delta_2=0.4$,~$\omega=1$,~$g_1=g2$,~$0~\leq~
g=g_1+g_2~\leq~3$. $E_{+}$ and $E_{-}$ are solutions with even and
odd parity respectively.\label{figure1}}
\end{figure}

This is the condition for an odd parity solution with photon number
bounded from above at $N=2$, coinciding with \cite{pj2}, which
depends on $\Delta_1$, $\Delta_2$ and $g$. So now, we have found the
algebraic structure and quasi-exact solutions with finite photon
numbers $N$. Furthermore, it is very interesting for the solution
with $N=1$, whose existing condition is independent of $g$. The
closed subspace is formed by
$\{|0,e,g\rangle,|0,g,e\rangle,|1,g,g\rangle,|1,e,e\rangle\}$, and
the condition is
\begin{eqnarray}\small
\det \left|
  \begin{array}{cccc}
    \Delta_1-\Delta_2-1 & 0                     & g/2                     & g/2                  \\
    0                   & \Delta_2-\Delta_1-1   & g/2                     & g/2                  \\
    g/2                 & g/2                   & -\Delta_1-\Delta_2      & 0                    \\
    g/2                 & g/2                   & 0                       & \Delta_2+\Delta_1    \\
   \end{array}
   \right|=0,
   \end{eqnarray}
which gives
\begin{equation}
(\Delta_1+\Delta_2)^2[(\Delta_1-\Delta_2)^2-1]=0,
\end{equation}
which is independent of $g$, coinciding with \cite{pj2}. So for
$\Delta_1-\Delta_2=1=\hbar\omega$ and
$\Delta_1-\Delta_2=-1=-\hbar\omega$, we obtain two quasi-exact
solutions
\begin{eqnarray}
|\psi\rangle_{g1}=\frac{1}{{\cal
N}}\left(\frac{2(\Delta_1+\Delta_2)}{g}|0,e,g\rangle+
 |1,g,g\rangle
-|1,e,e\rangle\right),\label{dk2}\\
 |\psi\rangle_{g2}=\frac{1}{{\cal
N}}\left(\frac{2(\Delta_1+\Delta_2)}{g}|0,g,e\rangle+
 |1,g,g\rangle
-|1,e,e\rangle\right),\label{dk3}
\end{eqnarray}
respectively, where ${\cal N}=\sqrt{4(\Delta_1+\Delta_2)^2+2g^2}/g$.
For example, choosing $\Delta_1=1.4$, $\Delta_2=0.4$, $g_1=g_2$, the
numerical spectrum of the two-qubit quantum Rabi model is shown in
figure \ref{figure1}. The horizontal line at $E=1=\hbar\omega$
corresponds to the special eigenstate $|\psi\rangle_{g1}$ (equation
\eqref{dk2}). This eigenstate exists in the whole coupling regime
with constant eigenenergy, like ``dark states'', but are coupling
dependent, and with at most $1$ photon.

For even parity, similarly, we obtain one such special eigenstate
\begin{eqnarray}
|\psi\rangle_{e}=\frac{1}{{\cal
N}}\left(\frac{2(\Delta_1-\Delta_2)}{g}|0,e,e\rangle-
 |1,e,g\rangle
+|1,g,e\rangle\right),\label{dk1}
\end{eqnarray}
 with the condition $\Delta_1+\Delta_2=1=\omega$, $g_1=g_2$ and
 $E=1=\hbar\omega$, consistent with \cite{pj2}. Now, we have demonstrated all the exceptional eigenstates of the
two-qubit quantum Rabi model with finite photon
 numbers presented in \cite{pj2} by finding its algebraic structure
 in the photon number space.

The special eigenstates $|\psi\rangle_{g1}$, $|\psi\rangle_{g2}$ and
$|\psi\rangle_{e}$ originate from the permutation symmetry of the
qubit-photon coupling terms, and we may conjecture there are similar
solutions for similar models. In the weak-coupling regime, the Rabi
model can reduce to Jaynes-Cummings model by the rotating-wave
approximation, so if there are similar special eigenstates for the
two-qubit Jaynes-Cunmmings model, we may find its application in a
simpler way. Now we try to find similar structure for the two-qubit
Jaynes-Cunmmings model \cite{2jc}
\begin{equation}
H_{tjc}=a^\dagger a+g_1(\sigma_1^{+}
a+\sigma_1^{-}a^\dagger)+g_2(\sigma_2^{+}
a+\sigma_2^{-}a^\dagger)+\Delta_1\sigma_{1z}+\Delta_2\sigma_{2z}.
\end{equation}
It is easy to find $C=a^\dagger
a+\frac{1}{2}(\sigma_{1z}+\sigma_{2z}+2)$ commutes with $H_{tjc}$,
so there is a conserved quantity $C$. Interestingly,
$|\psi\rangle_{e}$ (equation \eqref{dk1}) has a conserved quantity
$C=2$, and it is easy to testify $|\psi\rangle_{e}$ is also an
eigenstate of $H_{tjc}$ existing in the whole coupling regime with
constant eigenenergy for $g_1=g_2$ and $\Delta_1+\Delta_2=1$. To
find out all such kinds of eigenstates, we study the eigenproblem of
$H_{tjc}$. For $C=N$ ($N>1$), the Hamiltonian in the subspace
$\{|N-2,e,e\rangle,|N-1,e,g\rangle,|N-1,g,e\rangle,|N,g,g\rangle\}$
reads
\begin{eqnarray}\small\fl
\left(\begin{array}{cccc}
        N-2+\Delta_1+\Delta_2     &\sqrt{N-1} g_2                  & \sqrt{N-1}g_1               & 0 \\
        \sqrt{N-1}g_2             & N-1+\Delta_1-\Delta_2          & 0                           & \sqrt{N}g_1 \\
        \sqrt{N-1}g_1             & 0                              & N-1-\Delta_1+\Delta_2       & \sqrt{N}g_2 \\
        0                         & \sqrt{N}g_1                    & \sqrt{N}g_2                 & N-\Delta_1-\Delta_2
      \end{array}\right).
\end{eqnarray}
Using the time-independent Sch\"{o}dinger equation, we find the
eigenvalues $E$ is determined by
\begin{eqnarray}\label{en}\fl
(&E -N+\Delta_1+\Delta_2)(E -N+2-\Delta_1-\Delta_2) [(E -N+1)^2  -
(\Delta_1-\Delta_2)^2]\nonumber\\\fl &+(g_1^2+g_2^2)[(E-N+1)(E
-N+\Delta_1+\Delta_2)-2N(E-N+1)^2]\nonumber\\\fl
&+(g_1^2-g_2^2)[(g_1^2-g_2^2)(N^2-N)+(\Delta_1^2-\Delta_2^2)(2N-1)+(E+N)(\Delta_2-\Delta_1)]=0.
\end{eqnarray}
The condition (equation \eqref{en}) is generally dependent on $g_1$
and $g_2$, but there are two special cases. The first is the famous
``dark state''
$|\psi\rangle=\frac{1}{\sqrt{2}}(|N-1,e,g\rangle-|N-1,g,e\rangle$,
with the condition $g_1=g_2$ and $\Delta_1=\Delta_2$. The spin
singlet is decoupled from the photon field, so the eigenenergy and
eigenstate are coupling-independent. The second case is partly like
``dark state''---the eigenenergy is also coupling independent, but
the eigenstate is not. For $g_1=g_2$ and $\Delta_1+\Delta_2=1$,
equation \eqref{en} reduces to
\begin{eqnarray}
(E -N+1)^2 [(E -N+1)^2 + (\frac{1}{2}-N)g^2-
(\Delta_1-\Delta_2)^2]=0,
\end{eqnarray}
where $g=g_1+g_2$. For $E=N-1$, the condition is $g$-independent.
Besides, the eigenenergies are symmetric about $E=N-1$ and there are
two degenerate eigenstates with $E=N-1$ existing in the whole
coupling regime
\begin{eqnarray}\fl
|\psi_{C=N_a}\rangle=&\frac{1}{{\cal
A}}\left(\frac{2(\Delta_1-\Delta_2)}{\sqrt{N-1}g}|N-2,e,e\rangle-
 |N-1,e,g\rangle
+|N-1,g,e\rangle\right),\\\fl |\psi_{C=N_b}\rangle=&\frac{1}{{\cal
B}}\left(\frac{\sqrt{N-1}g}{(\Delta_1-\Delta_2)}|N-2,e,e\rangle+
 |N-1,e,g\rangle
-|N-1,g,e\rangle\right.\nonumber\\&\left.+\frac{(N-1)g^2+2(\Delta_1-\Delta_2)^2}{\sqrt{N}g(\Delta_2-\Delta_1)}|N,g,g\rangle\right),
\end{eqnarray}
where  $\frac{1}{{\cal A}}$ and $\frac{1}{{\cal B}}$ are the
normalizing constants. For $N=2$, $|\psi_{C=2_a}\rangle$ is just
$|\psi\rangle_{e}$ (equation \eqref{dk3}), which is the eigenstate
of the two-qubit quantum Rabi model.

For $C=1$, the subspace is formed by
$\{|0,e,g\rangle,|0,g,e\rangle,|1,g,g\rangle\}$, and the eigenvalues
satisfy
\begin{eqnarray}\label{c0}
E&[(\Delta_1-\Delta_2)^2+E(1-\Delta_1-\Delta_2)-E^2+g_1^2+g_2^2]\nonumber\\&+(\Delta_1+\Delta_2-1)(\Delta_1-\Delta_2)^2+(g_1^2-g_2^2)(\Delta_1-\Delta_2)=0.
\end{eqnarray}
For $g_1=g_2$ and $\Delta_1+\Delta_2=1$, equation \eqref{c0} reduces
to
\begin{equation}
 E[E^2-\frac{1}{2}g^2-(\Delta_1-\Delta_2)^2]=0.
\end{equation}
So there is an eigenstate existing in the whole coupling regime
with constant eigenenergy $E=0$
\begin{eqnarray}
|\psi_{C=0}\rangle=\frac{1}{{\cal
N}}\left(\frac{2(\Delta_1-\Delta_2)}{g}|1,g,g\rangle-
 |0,e,g\rangle
+|0,g,e\rangle\right).
\end{eqnarray}
For $C=0$, the eigenstate is $|0,g,g\rangle$, with constant
eigenenergy $E=-1$.

\begin{figure}[htbp]
\center
  \includegraphics[width=0.8\textwidth]{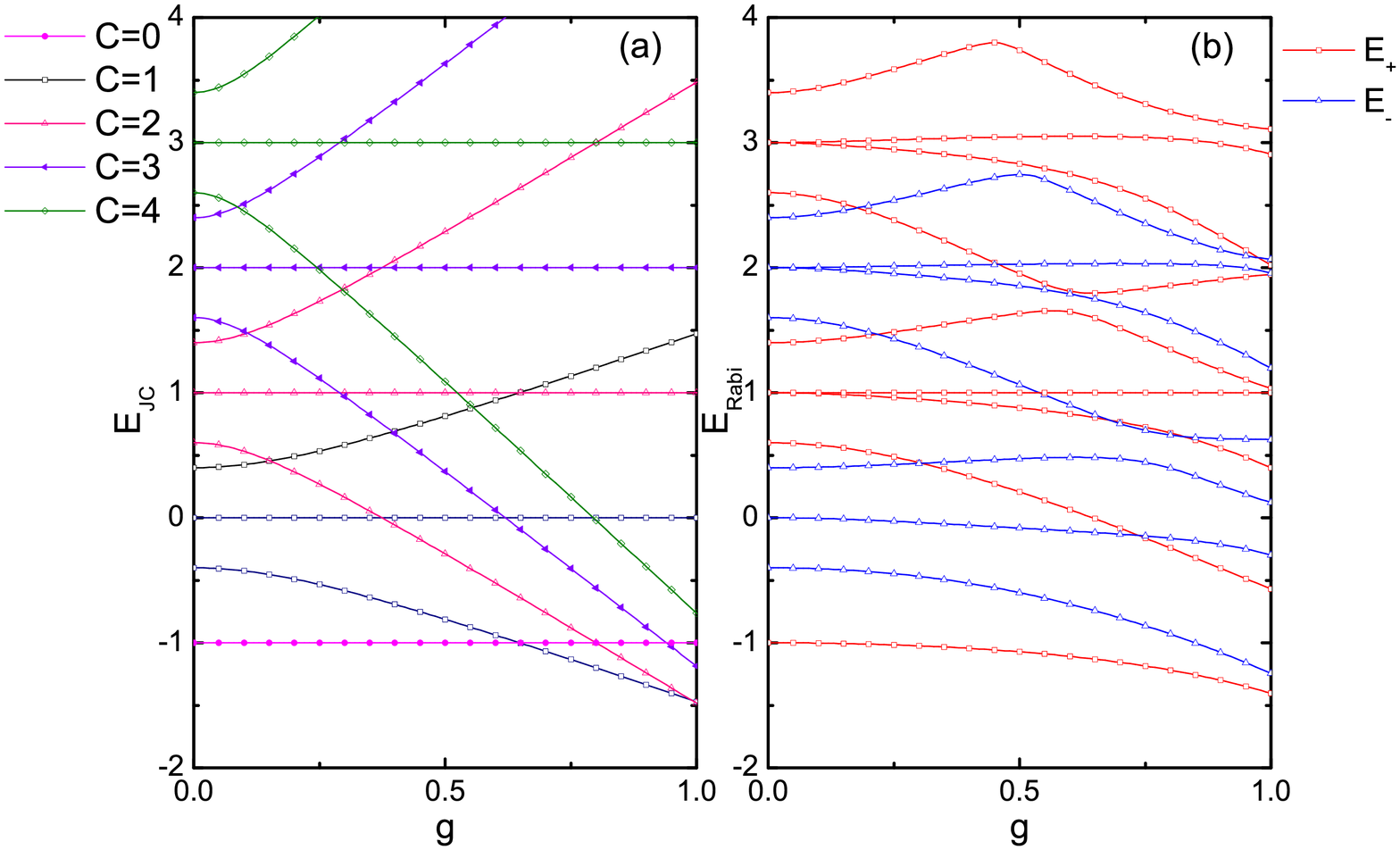}
\renewcommand\figurename{\textbf{Figure}}
\caption[2]{(a)~The spectrum of two-qubit Jaynes-Cummings model with
$\Delta_1=0.7$,~$\Delta_2=0.3$,~$\omega=1$,~$g_1=g_2$,~$0~\leq~
g=g_1+g_2~\leq~1$.~(b)~The numerical spectrum of two-qubit quantum
Rabi model with the same parameters. $E_{+}$ and $E_{-}$ are
solutions with even and odd parity respectively.\label{figure2}}
\end{figure}

To conclude, for identical-coupling $g_1=g_2$ and quasi-resonant
condition $\Delta_1+\Delta_2=1=\omega$, the spectrum of the
two-qubit Jaynes-Cummings Hamiltonian $H_{tjc}$ is very regular and
interesting: there are horizontal lines at $E=N$
($N=-1,0,1,\ldots$), and the energy curve with the same $C=N$ are
symmetric about the line $E=N-1$. For $C=0,1$, there is one kind of
eigenstates existing in the whole coupling regime with constant
eigenenergy, while for other cases, there are two such kinds of
degenerate eigenstaes, one of which for $C=2$ is also the eigenstate
of the two-qubit Rabi model. With constant eigenenergy, these
eigenstates are partly like ``dark state'', but they are coupling
dependent. Choosing $\Delta_1=0.7$, $\Delta_2=0.3$ and
$g_1=g_2=g/2$, the spectra of the two-qubit Jaynes-Cummings model
and Rabi model are compared in figure \ref{figure2}.

\section{Solvability of the two-qubit quantum Rabi model using Bogoliubov
operators} \label{s3} First for convenient, we make unitary
transformations $S_1=\frac{1}{\sqrt{2}}(\sigma_{1x}+\sigma_{1z})$
and $S_2=\frac{1}{\sqrt{2}}(\sigma_{2x}+\sigma_{2z})$ to the
two-qubit Rabi Hamiltonian (equation \eqref{gq}) to obtain ($\omega$
is set to 1)
\begin{equation}
H^{\prime}_{tq}=a^{\dagger}a+g_{1}\sigma_{1z}(a+a^{\dagger})+g_{2}\sigma_{2z}(a+a^{\dagger})
+\Delta_1\sigma_{1x}+\Delta_2\sigma_{2x}.
\end{equation}
$H^{\prime}_{tq}$ has a conserved parity with the $\mathbb{Z}_2$
transformation $R=T\otimes\sigma_{1x}\otimes\sigma_{2x}$, where
$T=\exp(i\pi a^\dagger a)$, giving us a way to diagonalize the
Hamiltonian in the basis of $\{|e\rangle_1,|g\rangle_1\}$, which is
the eigenvector of $\sigma_{1z}$. Applying the Fulton-Gouterman
transformation \cite{pj,fl},
\begin{eqnarray}
U=\left(\begin{array}{cc}
1&1\\
T\otimes\sigma_{2x}&-T\otimes\sigma_{2x}
\end{array}\right),
\end{eqnarray}
we obtain
\begin{eqnarray}
U^\dagger H^\prime_{tq} U=\left(
\begin{array}{cc}
H_+ &0 \\
0 & H_- \\
\end{array}
\right),
\end{eqnarray} where
\begin{equation}
H_{\pm}=a^\dagger
a+g_{1}(a+a^\dagger)+g_{2}(a+a^\dagger)\sigma_{2z}+\Delta_2\sigma_{2x}\pm\Delta_1
T\sigma_{2x},\label{+}
\end{equation}
acting on the subspace of $R$ with eigenvalues $\pm 1$. First we
consider $H_+$. For $H_-$, we just need to substitute $-\Delta_1$
for $\Delta_1$. In the basis of
$\{|e\rangle_2\otimes|\phi_1\rangle,|g\rangle_2\otimes|\phi_2\rangle\}$,
where $|\phi_1\rangle$ and $|\phi_2\rangle$ are photon field states,
$H_+$ is expanded as
\begin{eqnarray}
\left(\begin{array}{cc}
a^\dagger a+g(a+a^\dagger)&\Delta_2+\Delta_1 T\\
\Delta_2+\Delta_1 T&a^\dagger a+g^\prime(a+a^\dagger)
\end{array}\right).
\end{eqnarray}
To remove the linear terms of $a^\dagger$ and $a$, we use the
following Bogoliubov operators
\begin{equation}
A=a+g,~B=a+g^\prime.
\end{equation}
Firstly we use the Bogoliubov operator $A$. The time-independent
sch\"{o}rdinger equation reads
\begin{equation}
(A^\dagger
A-g^2-E)|\phi_1\rangle+\Delta_2|\phi_2\rangle+\Delta_1|\phi_4\rangle=0,\label{1}
\end{equation}
\begin{equation} [A^\dagger
A+(g^\prime-g)(A+A^\dagger)+g^2-2gg^\prime-E]|\phi_2\rangle+\Delta_2|\phi_1\rangle+\Delta_1|\phi_3\rangle=0,\label{2}
\end{equation}
where $|\phi_3\rangle=T|\phi_1\rangle$,
$|\phi_4\rangle=T|\phi_2\rangle$. To apply the reflection symmetry,
we make the transformation $T$ to \eqref{1} and \eqref{2} to obtain
\begin{eqnarray}
(A^{\dagger}
A-2g(A+A^{\dagger})+3g^2-E)|\phi_3\rangle+\Delta_2|\phi_4\rangle+\Delta_1|\phi_2\rangle=0,\\\label{3}
[A^{\dagger}
A-(g^\prime+g)(A+A^{\dagger})+g^2+2gg^\prime-E]|\phi_4\rangle+\Delta_2|\phi_3\rangle+\Delta_1|\phi_1\rangle=0.\label{4}
\end{eqnarray}
We expand the photon field states $|\phi_j\rangle,~j=1,~\ldots,~4$
in terms of the normalized orthogonal extended coherent state
\cite{qing}
\begin{eqnarray}
|n,g\rangle=\frac{e^{-g^2/2-g a^\dagger}}{\sqrt{n!}}(a^\dagger+g)^n,
\end{eqnarray}
which is the eigenstate of $A^\dagger A$, and obtain
\begin{equation}\label{5}
|\phi_j\rangle=e^{g^2/2}\sum_{n=0}^{\infty}\sqrt{n!}a_{j,n}|n,g\rangle,~j=1,\ldots,4.
\end{equation}
Substituting \eqref{5} into \eqref{1}--\eqref{4}, and left multiply
$\langle m,g|$, we obtain the recurrence relations for $a_{j,m}$
\begin{eqnarray}
(E-m+g^2)a_{1,m}=\Delta_1&a_{4,m}+\Delta_2a_{2,m},\label{ac1}\\
(g^\prime-g)(m+1)a_{2,m+1}=&(E-m+2g
g^\prime-g^2)a_{2,m}-(g^\prime-g)a_{2,m-1}\nonumber\\&-\Delta_1a_{3,m}-\Delta_2a_{1,m},\label{ac2}\\
2g(m+1)a_{3,m+1}=(m-&E+3 g^2)a_{3,m}-2g
a_{3,m-1}+\Delta_1a_{2,m}+\Delta_2a_{4,m},\label{ac3}\\
(g+g^\prime)(m+1)a_{4,m+1}=&(m-E+2g
g^\prime+g^2)a_{4,m}-(g+g^\prime)a_{4,m-1}\nonumber\\&+\Delta_1a_{1,m}+\Delta_2a_{3,m}.\label{ac4}
\end{eqnarray}
It is seen the coefficients $a_{j,m}$ depend on three initial
conditions, which can be chosen as $\{a_{1,0},~a_{2,0},~a_{3,0}\}$.

Then we consider the Bogoliubov operator $B=a+g^\prime$. Now $H_+$
is given as
\begin{eqnarray}
\left(\begin{array}{cc}
B^\dagger B+(g-g^\prime)(B+B^\dagger)+(g^\prime)^2-2gg^\prime&\Delta_2+\Delta_1 T\\
\Delta_2+\Delta_1 T&B^\dagger B-g^\prime
\end{array}\right).
\end{eqnarray}
Applying transformation $T$ to the time-independent sch\"{o}dinger
equation, we obtain four equations similar to \eqref{1}--\eqref{4}
\begin{eqnarray}\fl
(B^\dagger
B+(g-g^\prime)(B+B^\dagger)+(g^\prime)^2-2gg^\prime-E)|\varphi_1\rangle+\Delta_2|\varphi_2\rangle+\Delta_1|\varphi_4\rangle=0,\label{1a}\\\fl
(B^\dagger
B-(g^\prime)^2-E)|\varphi_2\rangle+\Delta_2|\varphi_1\rangle+\Delta_1|\varphi_3\rangle=0,\label{2a}\\\fl
(B^\dagger
B-(g^\prime+g)(B+B^\dagger)+(g^\prime)^2+2gg^\prime-E)|\varphi_3\rangle+\Delta_2|\varphi_4\rangle+\Delta_1|\varphi_2\rangle=0,\label{3a}\\\fl
(B^\dagger
B-2g^\prime(B+B^\dagger)+3(g^\prime)^2-E)|\varphi_4\rangle+\Delta_2|\varphi_3\rangle+\Delta_1|\varphi_1\rangle=0.
\label{4a}
\end{eqnarray}
Expanding the photon field states as
$|\varphi_j\rangle=e^{(g^\prime)^2/2}\sum_{n=0}^{\infty}\sqrt{n!}b_{j,n}|n,g^\prime\rangle,~
j=1,\ldots,4$, where the normalized extended coherent state
$|n,g^\prime\rangle$ is the eigenstate of $B$, and left multiplying
$\langle m,g^\prime|$, we obtain the recurrence relations for
$b_{j,m}$
\begin{eqnarray}
(g-g^\prime)(m+1)b_{1,m+1}=(&E-m+2g
g^\prime-(g^\prime)^2)b_{1,m}\nonumber\\&+(g^\prime-g)b_{1,m-1}-\Delta_1 b_{4,m}-\Delta_2 b_{2,m},\label{ad1}\\
(E-m+(g^\prime)^2)b_{2,m}=\Delta_1&b_{3,m}+\Delta_2b_{1,m},\label{ad2}\\
(g+g^\prime)(m+1)b_{3,m+1}=(&m-E+2g
g^\prime+(g^\prime)^2)b_{3,m}\nonumber\\&-(g+g^\prime)b_{3,m-1}+\Delta_1b_{2,m}+\Delta_2b_{4,m},\label{ad3}\\
2g^\prime(m+1)b_{4,m+1}=(m-&E+3
(g^\prime)^2)b_{4,m}\nonumber\\&-2g^\prime
b_{4,m-1}+\Delta_1b_{1,m}+\Delta_2b_{3,m}.\label{ad4}
\end{eqnarray}
There are three initial conditions, which can be chosen as
$\{b_{1,0},~b_{2,0},~b_{4,0}\}$. To utilize the reflection symmetry
$|\phi_1\rangle=T|\phi_3\rangle$, $|\phi_2\rangle=T|\phi_4\rangle$,
finally, we expand the photon states in terms of the photon number
states as $|\psi_j\rangle=\sqrt{n!}c_{j,n}|n\rangle$, and obtain the
recurrence relations for $c_{j,m}$
\begin{eqnarray}
(m+1)gc_{1,m+1}=(E-m)c_{1,m}-gc_{1,m-1}-\Delta_2c_{2,m}-\Delta_1c_{4,m},\label{ae1}\\
(m+1)g^\prime c_{2,m+1}=(E-m)c_{2,m}-g^\prime c_{2,m-1}-\Delta_2c_{1,m}-\Delta_1c_{3,m},\label{ae2}\\
(m+1)gc_{3,m+1}=(m-E)c_{3,m}-gc_{3,m-1}+\Delta_2c_{4,m}+\Delta_1c_{2,m},\label{ae3}\\
(m+1)g^\prime c_{4,m+1}=(m-E)c_{4,m}-g^\prime
c_{4,m-1}+\Delta_2c_{3,m}+\Delta_1c_{1,m}.\label{ae4}
\end{eqnarray}
Considering $|\psi_1\rangle=T|\psi_3\rangle$,
$|\psi_2\rangle=T|\psi_4\rangle$, we obtain $c_{1,m}=(-1)^m
c_{3,m}$, $c_{2,m}=(-1)^m c_{4,m}$, so there are only two initial
conditions, which can be chosen as $c_{1,0}$ and $c_{2,0}$.

States $|\phi_j\rangle$, $|\varphi_j\rangle$ and $|\psi_j\rangle$ in
different representations should be only different by a constant
(here can be chosen as $1$) if they are nondegenerate eigenstates
with eigenvalue $E$, so we obtain $8$ equations
\begin{eqnarray}
|\phi_j\rangle=|\varphi_j\rangle,\label{phi1}\\
|\varphi_j\rangle=|\psi_j\rangle.\label{phi2}
\end{eqnarray}

For practical calculation, we left multiply $\langle 0|e^{\beta}a$,
where $\beta$ is chosen arbitrarily, then \eqref{phi1} and
\eqref{phi2} are mapped to
\begin{eqnarray}
\langle 0|e^{\beta_1
a}|\phi_j\rangle&=\sum_{m=0}^{\infty}a_{j,m}\exp(-g\beta_1)(\beta_1+g)^m\nonumber\\
&=\langle 0|e^{\beta_1 a}|\psi_j\rangle=
\sum_{m=0}^{\infty}b_{j,m}\exp(-g^\prime\beta_1)(\beta_1+g^\prime)^m,\label{map1}\\
\langle 0|e^{\beta_2
a}|\varphi_j\rangle&=\sum_{m=0}^{\infty}b_{j,m}\exp(-g^\prime\beta_2)(\beta_2+g^\prime)^m\nonumber\\
&= \langle 0|e^{\beta_2
a}|\psi_j\rangle=\sum_{m=0}^{\infty}c_{j,m}\beta_2^m.\label{map2}
\end{eqnarray}

\begin{figure}[htbp]
\center
\includegraphics[width=0.8\textwidth]{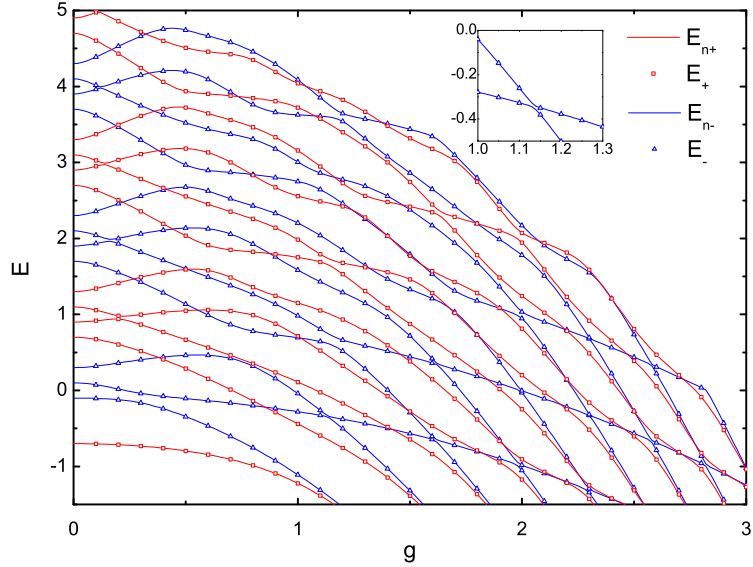}
\renewcommand\figurename{\textbf{Figure}}
\caption[2]{The spectrum of two-qubit quantum Rabi model with
$\Delta_1=0.4$,~$\Delta_2=0.3$,~$\omega=1$,~$g_1=3g_2$,~$0\leq
g=g_1+g_2\leq3$. $E_{n+}$ and $E_{n-}$ are numerical solutions with
even and odd parity respectively, while $E_{+}$ and $E_{-}$ are
analytical solutions with even and odd parity
respectively.\label{figure3}}
\end{figure}

Now we are still dealing with power series with infinite terms, so
to obtain clear reliable result, we must make all the power series
convergent. According to the recurrence relations for $a_{j,m}$
(equations \eqref{ac1}--\eqref{ac4}), $b_{j,m}$ (equations
\eqref{ad1}--\eqref{ad4}) and $c_{j,m}$ (equations
\eqref{ae1}--\eqref{ae4}), we find the radii of convergence of
corresponding power series are $|g-g^\prime|$,
$\min\{g-g^\prime,2g^\prime\}$ and $g^\prime$ respectively. So, for
different $g$ and $g^\prime$, we can always choose proper $\beta_1$
and $\beta_2$ to obtain convergent power series \cite{pj2}, so that
finite terms can give reliable results and by choosing proper
cutoff, and we can obtain the results with arbitrary accuracy. That
is the advantage of choosing these three different representations.
Because of the linearity of recurrence relations, we can denote
\begin{eqnarray}
\phi_j(\beta_1)=\langle 0|e^{\beta_1 a}|\phi_j\rangle=\sum_{k=1}^3
a_{k,0}\phi_j^k(\beta_1),\\
\varphi_j(\beta_1)=\langle 0|e^{\beta_1
a}|\varphi_j\rangle=\sum_{k=1,2,4}b_{k,0}\varphi_j^k(\beta_1),\\
\varphi_j(\beta_2)=\langle 0|e^{\beta_2
a}|\varphi_j\rangle=\sum_{k=1,2,4}b_{k,0}\varphi_j^k(\beta_2),\\
\psi_j(\beta_2)=\langle 0|e^{\beta_2 a}|\psi_j\rangle=\sum_{k=1}^2
c_{k,0}\psi_j^k(\beta_2),
\end{eqnarray}
where for example, $\varphi_j^k(\beta_1)$ is obtained by setting
$b_{k,0}$ equal to $1$ and other initial conditions equal to $0$ in
equations \eqref{ad1}--\eqref{ad4}, like in \cite{braak2}. Now we
have eight initial conditions for eight equations
\begin{eqnarray}
\phi_j(\beta_1)=\varphi_j(\beta_1),\\
\varphi_j(\beta_2)=\psi_j(\beta_2),
\end{eqnarray}
which can be denoted as
\begin{equation}
 M_{jk}e_k=0,
 \end{equation}
 with
 $\vec{e}=\{b_{1,0},b_{2,0},b_{4,0},a_{1,0},a_{2,0},a_{3,0},c_{1,0},c_{2,0}\}^T$.
 The determinant of $M$, which is just the function of energy $E$ must equal to
 $0$, so we obtain
\begin{equation}\label{gg}
G_+(E)=det(M_+)=0,
\end{equation}
which can be used to determine the eigenenergy $E$. Equation
\eqref{gg} takes similar form as equation (14) in \cite{pj2}, but
are obtained in a simper and more physical way. Choosing
$\Delta_1=0.4$, $\Delta_2=0.3$, $\omega=1$, $g_1=3g_2$, to have
convergent power series in equations \eqref{map1} and \eqref{map2},
we can choose $\beta_1=-3g_2$ and $\beta_2=-g_2$, then the results
can be obtained with arbitrary accuracy. The spectrum is shown in
figure \ref{figure3}. It is seen there are no level crossings within
the same parity subspace, so we can label each eigenstates with two
quantum numbers---energy level and parity, but the total degrees of
freedom are three, so according to the quantum integrability
criterion proposed by Braak \cite{br}, the model is non-integrable,
consistent with what the narrow avoided crossings in the same parity
subspace indicate \cite{xu} and the result in \cite{pj2}.

\section{Conclusions}\label{s4}
We have clarified the algebraic structure behind the possibility of
the quasi-exact solutions with finite photon numbers found in
\cite{pj2}. By analyzing the Hamiltonian structure in the photon
number space, we find that the permutation symmetry of the
qubit-photon coupling terms for the two qubits brings about closed
subspace, and hence quasi-exact solutions for certain parameters.
The novel coupling-dependent eigenstates existing in the whole
coupling regime with constant eigenenergy $E$ equal to single photon
energy $\hbar\omega$ correspond to quasi-exact solutions with at
most $1$ photon, with the condition for the qubits energy splittings
$\Delta_1\pm\Delta_2=\hbar\omega$ or
$\Delta_2-\Delta_1=\hbar\omega$. We have demonstrated this directly
from the Hamiltonian structure. These special eigenstates are partly
like ``dark states'', but are coupling-dependent, which may have
some potential application. Furthermore, based on our study on the
two-qubit quantum Rabi model, we conjecture such ``dark
states''-like eigenstates commonly exist in similar models with
permutation symmetry of the qubit-photon coupling terms. For
example, for the homogenous coupled two-qubit Jaynes-Cummings model,
there are many such kinds of eigenstates with constant energy
$E=N\hbar\omega~(N=-1,0,1,\ldots)$ in the whole coupling regime,
with the condition $\Delta_1+\Delta_2=\hbar\omega$. One of these
special states is also the eigenstate of the two-qubit quantum Rabi
model. Since the Jaynes-Cummings model is simper than the Rabi
model, we may find the application of these special eigenstates
easier.

Besides, using Bogoliubov operators, we have analytically retrieved
the solution of the two-qubit quantum Rabi model. We find three
different representations to expand the Hamiltonian, and the
solutions can be determined by convergent power series. In this way,
the eigenproblem of the infinite dimensional Hamiltonian can reduces
to finite dimensional in practical calculation reasonably, and the
results can reach arbitrary accuracy. Without using Bargmann space,
this method is more physical and concise.

\section*{Acknowledgements}
JP is thankful to Yibin Qian and Daniel Braak for helpful
discussions. This work was supported by the National Natural Science
Foundation of China (Grants Nos 11347112, 11204263, 11035001,
11404274, 10735010, 10975072, 11375086 and 11120101005), by the 973
National Major State Basic Research and Development of China (Grants
Nos 2010CB327803 and 2013CB834400), by CAS Knowledge Innovation
Project No. KJCX2-SW-N02, by Research Fund of Doctoral Point (RFDP)
Grant No. 20100091110028, by the Project Funded by the Priority
Academic Program Development of Jiangsu Higher Education
Institutions (PAPD), by the Scientific Research Fund of Hunan
Provincial Education Department (No. 12C0416).

%
%

\end{document}